\documentclass[epj,nopacs]{svjour}

\usepackage{latexsym}
\usepackage[numbers,sort&compress]{natbib}
\usepackage{graphics}
\usepackage{graphicx}
\usepackage{subfigure}
\usepackage{hyperref}
\usepackage[tbtags]{amsmath}
\usepackage{color}

\begin{document}

\title{Extended Birkhoff's Theorem in the $f(T)$ Gravity}
\author{Han Dong \inst{1} \thanks{\email{donghan@mail.nankai.edu.cn}},
Ying-bin Wang \inst{1} \thanks{\email{wybn@mail.nankai.edu.cn}}
\and Xin-he Meng\inst{1,2} \thanks{\email{xhm@nankai.edu.cn}(corresponding author)}
}

\institute{Department of physics, Nankai University, Tianjin 300071, China
\and Kavli Institute of Theoretical Physics China, CAS, Beijing 100190, China}

\date{Received: date / Revised version: date}

\abstract{ The $f(T)$ theory, a generally modified teleparallel gravity, has
been proposed as an alternative gravity model to account for the
dark energy phenomena. Following our previous work [Xin-he Meng and Ying-bin Wang, EPJC(2011), arXiv:1107.0629v1],
we prove that the Birkhoff's theorem holds in a more general context,
specifically with the off diagonal tetrad case, in this communication letter.
Then, we discuss respectively the results of the external vacuum and
internal gravitational field in the $f(T)$ gravity framework,
as well as the extended meaning of this theorem.
We also investigate the validity of the Birkhoff's theorem in the frame of $f(T)$ gravity
via conformal transformation by regarding the Brans-Dicke-like scalar as effective matter,
and study the equivalence between both Einstein frame and Jordan frame.
\\\\
PACS.98.80.Cq Modified theories of gravity } \maketitle

\section{Introduction}

Since the discovery of the accelerating expansion of the universe, people
have made great efforts to investigate the hidden mechanism, which
also provides us with great opportunities to deeply probe the fundamental theories of
gravity. As one of modified gravitational theories, the
$f(T)$ gravity is firstly invoked to drive inflation by Ferraro and Fiorini \cite{fT0}. Later, Bengochea and
Ferraro \cite{fT}, as well as Linder \cite{fT1}, propose to use the $f(T)$ theory to drive the current
accelerated expansion of our universe without invoking the mysterious dark energy. The framework is a
generalization of the so-called \textit{Teleparallel Equivalent of
General Relativity} (TEGR) which is firstly propounded by Einstein in
1928 \cite{einstein} and maturates in the 1960s (For some reviews,
see \cite{TEGR1,TEGR2}). Contrary to the theory of general
relativity which is based on Riemann geometry involving only
curvature, the TEGR is based on the so named Weitzenb\"ock geometry
with the non-vanishing torsion. Owing to the definition of
Weitzenb\"ock connection rather than the Levi-Civita connection, the
Riemann curvature is automatically vanishing in the TEGR framework,
which brings the theory a new name, \textit{Teleparallel Gravity}.
For a specific choice of parameters, the TEGR behaves completely equivalent to the Einstein's theory of general
relativity. Furthermore, by using the torsion scalar $T$ as the
Lagrangian density, the TEGR can give a field equation with second
order only, instead of the fourth order as in the Einstein's field equation,
and avoids the instability problems caused by higher order derivatives
as demonstrated in the metric framework $f(R)$ gravity models.

Similar to the generalization of Einstein's theory of general relativity to the
$f(R)$ theory (For some references,
see \cite{fR0,fRrev,fR1,fR11,fRa0,fRa1,fRa2,fRa3,fRa4,fRa5,fRa6,fRa7,fRa8,fRa9,fR2,fR3,fR4}),
the modified version of teleparallel gravity assumes a general function
$f(T)$ as the model Lagrangian density. Also, the
$f(T)$ theory can be directly reduced to the TEGR if we choose the
simplest case, that is, $f(T)=T$. The Lorentz invariance and conformal invariance
of the $f(T)$ theory is also investigated \cite{fT_Lorentz,fT_conformal},
with many interesting results presented. A class of $f(T)$ models with diagonal tetrad are
proposed in succession to explain the late-time acceleration of the
cosmic expansion without the mysteriously so-called dark energy, and
are fitted the cosmological data-sets very well (e.g.
\cite{fT,fT1,fT_w,fT2,fT3,fT4,fT5,fT6,fT7}). Most of the previous works
consider the $f(T)$ gravity with diagonal
tetrad field only. Noting that the tetrad field has sixteen components
rather than ten as in the metric frame, there are more freedoms
and more physical meaning from the extra uncertain six components.
In our previous work \cite{fT_birkhoff}, we have proved the validity of Birkhoff's
theorem in $f(T)$ gravity with a specific diagonal tetrad.
In this letter, we study this issue more generally with also the off diagonal tetrad field, and discuss the
physical meaning in a more extended context.

The Birkhoff's theorem is also called Jebsen-Birkhoff theorem,
for it was actually discovered by Jebsen two years before George D. Birkhoff
in 1923 \cite{birkhoff,bb}. The theorem states that the spherically
symmetric gravitational field in vacuum must be static, with a
metric uniquely given by the Schwarzschild solution form of Einstein
equations \cite{weinberg}. It is well known that the Schwarzschild
metric is found in 1918 as the external (vacuum) solution of a
static and spherical symmetric star. The Birkhoff's theorem means that any spherically
symmetric object possesses the same static gravitational
field, as if the mass of the object were concentrated at the center.
Even if the central spherical symmetric object is dynamic motion,
such as the case in the collapse and pulsation of stars, the external gravitational field is
still static if only the radial motion is spherically symmetric.
The same feature is held in the classical Newtonian gravity.

In this work we investigate the Birkhoff's theorem in the $f(T)$ gravity model
generally with both the diagonal and the off diagonal tetrad fields,
analyze the extended meaning of this theorem, and study the equivalence between both Einstein frame and Jordan frame.
First, in section two we briefly review the $f(T)$ theories,
and in section three we prove the validity of Birkhoff's theorem
of the $f(T)$ gravity with both off diagonal tetrad and diagonal tetrad fields.
In section four, we then discuss the validity of the Birkhoff's theorem in the frame of $f(T)$ gravity
via conformal transformation by regarding the Brans-Dicke-like scalar as effective matter.
Both the Jordan and Einstein frames are discussed in this section.
And some new conclusions and discussions are provided in the last section.

\section{Elements of $f(T)$ Gravity}

Instead of the metric tensor, the vierbein field
$\mathbf{e}_{i}(x^{\mu})$ plays the role of the dynamical variable in the
teleparallel gravity. It is defined as the orthonormal basis of the
tangent space at each point $x^{\mu}$ in the manifold, namely,
$\mathbf{e}_{i}\cdot \mathbf{e}_{j}=\eta_{ij}$, where
$\eta_{ij}=diag(1,-1,-1,-1)$ is the Minkowski metric. The vierbein
vector can be expanded in spacetime coordinate basis:
$\mathbf{e}_{i}=e^{\mu}_{i} \partial_{\mu}$,
$\mathbf{e}^i=e^i_\mu{\rm d}x^\mu$. According to the convention,
Latin indices and Greek indices, both running from 0 to 3, label the
tangent space coordinates and the spacetime coordinates
respectively. The components of vierbein are related by $e_{\mu}^i
e^{\mu}_j=\delta^{~i}_{j}$, ~~$e_{\mu}^i
e^{\nu}_i=\delta_{\mu}^{~\nu}$.

The metric tensor is determined uniquely by the vierbein as
\begin{equation}
g_{\mu\nu}=\eta_{ij} e_{\mu}^i e_{\nu}^i,
\end{equation}
which can be equivalently expressed as $\eta_{ij}=g_{\mu\nu}
e^i_{\mu} e^j_{\nu}$. The definition of torsion tensor is given then by
\begin{equation}
T^{\rho}_{~\mu\nu}=\Gamma^{\rho}_{~\nu\mu}-\Gamma^{\rho}_{~\mu\nu},
\end{equation}
where $\Gamma^{\rho}_{~\mu\nu}$ is the connection. Evidently,
$T^{\rho}_{~\mu\nu}$ vanishes in the Riemann geometry since the
Levi-Civita connection is symmetric with respect to the two
covariant indices. Differing from that in Einstein's theory of
general relativity, the teleparallel gravity uses Weitzenb\"ock
connection defined directly from the vierbein:
\begin{equation}
\Gamma^{\rho}_{~\mu\nu}=e_i^{\rho} \partial_{\nu} e^i_{\mu}.
\end{equation}
Accordingly, the antisymmetric non-vanishing torsion is
\begin{equation}\label{torsion}
T^{\rho}_{~\mu\nu}=e_i^{\rho}(\partial_{\mu}e^i_{\nu} - \partial_{\nu}e^i_{\mu}).
\end{equation}
It can be confirmed that the Riemann curvature in this framework is precisely vanishing:
\begin{equation}
R^\rho_{~\theta\mu\nu}=\partial_\mu \Gamma^\rho_{~\theta\nu}-\partial_\nu \Gamma^\rho_{~\theta\mu}+\Gamma^\rho_
{~\sigma\mu}\Gamma^\sigma_{~\theta\nu}-\Gamma^\rho_{~\sigma\nu} \Gamma^\sigma_{\theta\mu}=0.
\end{equation}

In order to get the action of the teleparallel gravity, it is
convenient to define other two tensors:
\begin{equation}\label{contorsion}
K^{\mu\nu}_{~~\rho}=-\frac{1}{2}(T^{\mu\nu}_{~~\rho}-T^{\nu\mu}_{~~\rho}-T_{\rho}^{~\mu\nu}),
\end{equation}
and
\begin{equation}\label{S}
S_\rho^{~\mu\nu}=\frac{1}{2}(K^{\mu\nu}_{~~\rho}+\delta_\rho^{~\mu}T^{\theta\nu}_{~~\theta}-\delta_\rho^{~\nu}T^
{\theta\mu}_{~~\theta}).
\end{equation}
Then the torsion scalar as the teleparallel Lagrangian density is defined by
\begin{equation}\label{T}
T=S_{\rho}^{~\mu\nu} T^{\rho}_{~\mu\nu}.
\end{equation}
The action of teleparallel gravity is then expressed as
\begin{equation}
I=\frac{1}{16\pi G}\int {\rm d}^4 x~e\,T ,
\end{equation}
where $e=$det$(e^i_{\mu})=\sqrt{-g}$. Performing variation of the
action with respect to the vierbein, one can get the equations of
motion which are equivalent to the results of Einstein's theory of
general relativity.

Just as in the $f(R)$ theory, the generalized version of
teleparallel gravity could be obtained by extending the Lagrangian
density directly to a general function of the scalar torsion $T$ :
\begin{equation}\label{action}
 I=\frac{1}{16\pi G}\int {\rm d}^4x~e\,f(T).
\end{equation}
This modification is expected possibly to provide a natural way to understand
the cosmological observations, especially for the dark energy
phenomena, as a motivation. Then the variation of the action with respect
to vierbein leads to the following equations:
\begin{equation}\label{field eqn}
\begin{split}
{\big[}e^{-1}e^i_\mu\partial_\sigma(eS_i^{~\sigma\nu})-T^\rho_{~\sigma\mu}S_\rho^{~\nu\sigma}{\big]}f_T+
S_\mu^{~\rho\nu}\partial_\rho Tf_{TT}\\
-\frac{1}{4}\delta_\mu^{~\nu}f=4\pi GT_\mu^{~\nu} ,
\end{split}
\end{equation}
where $f_T$ and $f_{TT}$ represent the first and second order
derivatives with respect to $T$ respectively, and
$S_i^{~\sigma\nu}=e_i^\rho S_\rho^{~\sigma\nu}$. ~$T_\mu^{~\nu}$ is
the energy-momentum tensor of the particular matter, with assuming
that matter couples to the metric in the standard form.

\section{Extended Birkhoff's theorem in $f(T)$ Gravity with both the off diagonal tetrad and the diagonal tetrad}
\label{sec3}
In our previous work \cite{fT_birkhoff} we have proved that the Birkhoff's theorem
is valid in $f(T)$ gravity with diagonal tetrad.
But the $f(T)$ gravity with diagonal tetrad will give a strong constraint
for a constant torsion scalar, which is shown in Ref. \cite{fT_relativistic stars},
while another off diagonal tetrad field is able to construct interesting exact solutions to these field equations.
The off diagonal tetrad field can provide bigger room to modify the gravity for its six more freedoms.
Therefore, we think that the research of the Birkhoff's theorem
with off diagonal tetrad field is necessary and realistically meaningful.

We consider the external vacuum gravitational field
solution of a spherically symmetric object. The spherically
symmetric metric generally can always be written in the following form:
\begin{equation}\label{metric}
{\rm d} s^2={\rm e}^{a(t,r)}{\mathrm{d}}t^2-{\rm e}^{b(t,r)}{\mathrm{d}}r^2-r^2{\mathrm{d}}\theta^2-r^2 \sin^2\!\theta~{\mathrm{d}}\phi^2,
\end{equation}
where $a(t,r)$, $b(t,r)$ are arbitrary functions of the coordinates $t$ and $r$.
One possible corresponding off diagonal tetrad field can be written as
\begin{equation}
e^i_\mu =
\left( \begin{array}{cccc}
e^{\frac{a(t,r)}{2}} & 0                                                & 0                       & 0     \\
0                    & {\rm e}^{\frac{b(t,r)}{2}}\!\sin\!\theta \cos\!\phi\, & r\!\cos\!\theta \cos\!\phi\, & -r\!\sin\!\theta \sin\!\phi  \\
0                    & {\rm e}^{\frac{b(t,r)}{2}}\!\sin\!\theta \sin\!\phi\, & r\!\cos\!\theta \sin\!\phi\, & -r\!\sin\!\theta \cos\!\phi  \\
0                    & {\rm e}^{\frac{b(t,r)}{2}}\!\cos\!\theta\,            & -r\!\sin\!\theta\,           & 0     \\
\end{array} \right)
\end{equation}
The determinant of vierbein is $e={\rm e}^{\frac{a(t,r)+b(t,r)}{2}}r^2\sin\!\theta$.
Then the tensors defined in Eqs. (\ref{torsion},\ref{contorsion},\ref{S}) are determined,
and the torsion scalar is given by
\begin{equation}\label{T2}
T=\frac{2{\big(}{\rm e}^{\frac{b(t,r)}{2}}-1{\big)}{\big(}{\rm e}^{\frac{b(t,r)}{2}}-a^\prime\!(t,r)\,r-1{\big)}}{{\rm e}^{b(t,r)}r^2},
\end{equation}
where a prime denotes the derivative with respect to $r$, while a dot overhead denotes
the derivative with respect to $t$. We will follow these conventions throughout this work.

For convenience, we introduce the tensor $E_\mu^{~\nu}$ to represent of
the left hand side of Eq. (\ref{field eqn}), and the field
equation can be then re-expressed concisely as
\begin{equation}
E_\mu^{~\nu}=4\pi GT_\mu^{~\nu}.
\end{equation}
Then, we work out all the components of
$E_\mu^{~\nu}$, and find most of them are not vanishing, including
some quite complicated ones. But the two components we used, fortunately
not very complex, are given by respectively
\begin{eqnarray}
E_2^{~0}&=&\frac{1}{4}\cot\!\theta \, e^{-a(t,r)} \, \dot b(t,r) f_T,\label{E20}\\
E_2^{~1}&=&\frac{1}{4 r} \cot\!\theta \, e^{-b(t,r)} \, {\big(}2-2{\rm e}^{\frac{b(t,r)}{2}}+a^\prime(t,r) r{\big)}f_T.\label{E21}
\end{eqnarray}
Since the non-diagonal elements of energy-momentum tensor for spherically symmetric
gravitational source are naturally equal to zero, $E_2^{~0}$ always vanishes.
And $f_T$ should not be trivially zero for the real universe observation, which is restricting $b(t,r)$
to be only the function of $r$, that is,
\begin{equation}\label{B}
b(t,r)=b(r).
\end{equation}
For the same reason as to $E_2^{~0}$, that $E_2^{~1}$ is also equal to zero. After some manipulations, Eq. (\ref{E21}) leads to
\begin{equation}\label{constrain}
1+\frac{a(t,r)^\prime r}{2}={\rm e}^{\frac{b(r)}{2}}.
\end{equation}
For $b(r)$ is independent of $t$, the left of Eq. (\ref{constrain}) should be also a function of $r$.
After performing an integration with respect to $r$, the function $a(t,r)$ could be simply expressed as
\begin{equation}\label{A}
a(t,r)=\widetilde a(r)+c(t),
\end{equation}
where $c(t)$ is an arbitrary function of $t$, and the $\widetilde a(r)$ is an integral function of the variable $r$.
Therefore the function ${\rm e}^{a(t,r)}$ can be written as
\begin{equation}
{\rm e}^{a(t,r)}={\rm e}^{\widetilde a(t)}{\rm e}^{c(t)}.
\end{equation}
The factor ${\rm e}^{c(t)}$ can always be absorbed in the metric through a coordinate
transformation $t\to t^\prime$, where $t^\prime$ is the new time coordinate defined as:
 \begin{equation}
 {\rm d}t^\prime={\rm e}^{\frac{c(t)}{2}} {\rm d}t.
 \end{equation}
Therefore the metric presented in Eq. (\ref{metric}) becomes
\begin{equation}\label{metric1}
{\rm d} s^2={\rm e}^{\widetilde a\!(r)}~{\mathrm{d}}t^2-{\rm e}^{b\!(r)}~{\mathrm{d}}r^2-r^2{\mathrm{d}}\theta^2-r^2 \sin^2\!\theta~{\mathrm{d}}\phi^2,
\end{equation}
This is exactly a static metric, which is required by the Birkhoff's theorem.

As a demonstration, we also perform the computation with the corresponding diagonal tetrad field, which can be written as
\begin{equation}
e^i_\mu={\rm diag}{\big(}{\rm e}^{\frac{a(t,r)}{2}},~{\rm e}^{\frac{b(t,r)}{2}},~r,~r\sin\!\theta{\big)},
\end{equation}
and the determinant of vierbein is $e={\rm e}^{\frac{a(t,r)+b(t,r)}{2}}r^2\sin\theta$.
It is easy to find some of the non-vanishing components of $E_\mu^{~\nu}$ .
The two components we used are given by respectively
\begin{equation}
E_1^{~0} = -\frac{{\rm e}^{-a(t,r)}}{2 r} \dot b(t,r) f_T,\label{e10}
\end{equation}
and
\begin{equation}
\begin{split}
E_0^{~2} = \frac{2{\rm e}^{- b(t,r)} \cos\theta}{r^4 \sin\theta} \left(\dot{a}^\prime(t,r) r-\dot{b}(t,r)-r \dot{b}(t,r) a^\prime(t,r)\right)\\
\cdot f_{TT}.\label{e02}
\end{split}
\end{equation}
Since the non-diagonal elements of energy-momentum tensor are naturally equal to zero,
$E_1^{~0}$ and $E_0^{~2}$ always vanishes. Like the previous discussion, $\dot b(t,r) = 0$, we get
\begin{equation}\label{b}
b(t,r)=b(r).
\end{equation}
Substituted into equation (\ref{e02}), it leads to
\begin{equation}\label{constrain2}
\dot{a}^\prime(t,r) r=0.
\end{equation}
Similar to Eqs.(\ref{B}) and (\ref{constrain}),
the above two equations deduce the same conclusion as the case of the off diagonal tetrad field.

In the previous proof, we have not used any specific model form for $f(T)$ gravity.
What we only require is that it should satisfy the necessary physical meaning, which means nontrivially $f_T \neq 0$.

Note that the integral performed on Eq.(\ref{constrain}) is over the external
region, and therefore the distribution and motion of the internal source matter cannot influence
$\widetilde a(r)$ any way. We then come to the conclusion that the spherically symmetric
vacuum solution of the $f(T)$ gravity must be static, and
is independent of the radial distribution and motion of the
source matter, implying that Birkhoff¡¯s theorem still holds
generally.

\section{The Birkhoff's theorem in the frame of $f(T)$ Gravity via conformal transformation}
It is well known that the $f(R)$ gravity is dynamically equivalent
to a particular class of scalar-tensor theories via conformal transformation,
but the Birkhoff's theorem generally does not hold in scalar-tensor gravity.
The case of $f(T)$ gravity via conformal transformation is more complicated
than that of $f(R)$ theories, which has been proved in the work \cite{fT_conformal}.

We will explore the difference between $f(T)$ gravity and scaler-tensor theory,
and compare the results obtained from the Jordan and Einstein frames via conformal transformation.
Firstly, let us write the general action for a Brans-Dicke-like $f(T)$ theory,
\begin{equation}\label{SBD}
    S_{BD} = \int {\rm d}^4 x~e\bigg[ \phi T - \frac{\omega}{\phi}g^{\mu\nu} \nabla_{\mu}\phi \nabla_{\nu}\phi
    - V(\phi) + 2k^2\mathcal{L}_m(e_{\mu}^{~i}) \bigg],
\end{equation}
where we have assumed $\omega$ to be constant. This action is written in the Jordan frame,
which is related with the Einstein frame by the conformal transformation,
\begin{equation}\label{trans}
    e_{\mu}^{~i}=\Omega^{-1} \tilde e_{\mu}^{~i}, \quad where \quad \phi=\Omega^{2}
\end{equation}
under which the action (\ref{SBD}) can be transformed to the Einstein frame as,
\begin{eqnarray}\label{SE1}
    S_{E} &=& \int {\rm d}^4 x~\tilde{e}\bigg[\tilde{T} - 2\phi^{-1}\tilde \partial^{\mu}\phi \tilde T^{\rho}_{~\rho\mu}
    - \frac{\omega-3/2}{\phi^2}\tilde \nabla_{\mu}\phi \tilde \nabla^{\mu}\phi \nonumber \\
&-& \frac{V(\phi)}{\phi^2} \bigg] + 2k^2 \int {\rm d}^4 x~\tilde{e}\,\mathcal{\tilde  L}_m(\tilde e_{\mu}^{~i})
\end{eqnarray}
By redefining the scalar field as $\phi=e^{\varphi/\sqrt{2\omega-3}}$
for the observation of the solar system $\omega\approx500$, and $U(\varphi)=\frac{V(\phi)}{\phi^2}$,
we change the action (\ref{SE1}) as,
\begin{eqnarray}\label{SE2}
    S_{E}
    &=& \int {\rm d}^4 x~\tilde{e}\bigg[\tilde{T}
    - \frac{2}{\sqrt{2\omega-3}}\tilde \partial^{\mu}\varphi \tilde T^{\rho}_{~\rho\mu}
    - \frac{1}{2}\tilde g^{\mu\nu}\tilde \nabla_{\mu}\varphi\tilde \nabla_{\nu}\varphi \nonumber \\
    &-& U(\varphi)\bigg] + 2k^2 \int {\rm d}^4 x~\tilde{e}\,\mathcal{\tilde L}_m(\tilde e_{\mu}^{~i})
\end{eqnarray}
Differing from the case in $f(R)$ gravity, an additional scalar-torsion coupling term presents in the action.
Therefore, the $f(T)$ gravity is not simply dynamically equivalent to the TEGR action
plus a scalar field via conformal transformation,
and one cannot use the results of scalar-tensor theories directly to $f(T)$ gravity.
Nonetheless, we can also obtain the field equations by varying the action (\ref{SE2}) with respect to the
tetrad field $e_{\alpha}^{~i}$ and the scalar field $\varphi$, which yields
\begin{eqnarray}
    \tilde e^{-1}\tilde G^{\alpha}_{~i}
    &=& \frac{1}{2\sqrt{2\omega-3}}\tilde\partial_{\mu}\big[\tilde\partial^{\alpha}\varphi \tilde e^{\mu}_{~i}
    - \tilde\partial^{\mu}\varphi \tilde e^{\alpha}_{~i}\big] \nonumber \\
    &-& \frac{\tilde\partial^{\mu}\varphi}{2\sqrt{2\omega-3}}\tilde e^{\alpha}_{~i}\tilde T^{\rho}_{~\rho\mu}
    + \frac{\tilde\partial^{\mu}\varphi}{2\sqrt{2\omega-3}}\tilde e^{\rho}_{~i}\tilde T^{\alpha}_{~\rho\mu}\nonumber \\
    &+& \frac{1}{4}\tilde e^{\nu}_{~i}\tilde \nabla^{\alpha}\varphi\tilde \nabla_{\nu}\varphi
    - \frac{1}{8}\tilde e^{\alpha}_{~i}\tilde \nabla^{\sigma}\varphi\tilde \nabla_{\sigma}\varphi
     \nonumber \\
    &-& \frac{1}{4}e^{\alpha}_{~i}U(\varphi) + \frac{k^2}{2}e^{\rho}_{~i}\tilde T^{\alpha \,(m)}_{~\rho} , \label{field eqn0}
    \end{eqnarray}
and
\begin{eqnarray}\label{scalar field eqn0}
    -2k^2\frac{\delta(\tilde{e}\,\mathcal{\tilde L}_m)}{\tilde{e}\delta\varphi} &=& \tilde \Box \varphi
    - \frac{{\rm d}U(\varphi)}{{\rm d}\varphi}  \nonumber \\
    &+& \frac{2}{\sqrt{2\omega-3}}\tilde e^{-1}\tilde\partial_{\mu}\big(\tilde{e}\tilde g^{\mu\nu}\tilde T^{\rho}_{~\rho\nu}\big),
\end{eqnarray}
where the $\tilde G^{\alpha}_{~i}$ in Eq.(\ref{field eqn0}) is defined by
\begin{equation}
    \tilde G^{\alpha}_{~i} = \tilde\partial_{\mu}(\tilde{e}\tilde e^{\rho}_{~i}\tilde S_{\rho}^{~\mu\alpha})
    + \tilde{e}\tilde e^{\nu}_{~i}\tilde T^{\rho}_{~\mu\nu}\tilde S_{\rho}^{~\mu\alpha}
    - \frac{1}{4}\tilde{e}\tilde e^{\alpha}_{~i}\tilde T.
\end{equation}
The field equation (\ref{field eqn0}) seems very complicated, while the components that we need can be simplified.
If the $\varphi=\varphi_{0}$, which is a constant, the field equation (\ref{field eqn0}) degenerates to the
teleparallel gravity with the cosmological constant $\Lambda = 2 U(\varphi)$.
In this case the Birkhoff's Theorem holds both in Einstein frame
and Jordan frame with the diagonal and off diagonal tetrad fields.

Assuming $\varphi=\varphi(t,r)$, because of the relation $\phi=e^{\varphi/\sqrt{2\omega-3}}$,
we can generally define $\phi=\phi(t,r)$.
On one hand, the $E_2^{~0}$ and $E_2^{~1}$, which we use to prove the validity
of Birkhoff's Theorem with the off diagonal tetrad field, change as,
\begin{eqnarray}
E_2^{~0}&=&\frac{1}{4}\cot\!\theta \, e^{-a(t,r)} \, \dot b(t,r) f_T
+ \frac{\tilde\partial^{\mu}\varphi}{2\sqrt{2\omega-3}}\tilde T^{t}_{~\theta\mu}, \\
E_2^{~1}&=&\frac{1}{4 r}\cot\!\theta \, e^{-b(t,r)} \, {\big(}2-2{\rm e}^{\frac{b(t,r)}{2}}+a^\prime(t,r) r{\big)}f_T \nonumber \\
&+& \frac{\tilde\partial^{\mu}\varphi}{2\sqrt{2\omega-3}}\tilde T^{r}_{~\theta\mu}.
\end{eqnarray}
Then we consider the non-zero components of the antisymmetric torsion tensor $T^{\rho}_{~\mu\nu}$ for
the two covariant indices with the off diagonal tetrad field,
\begin{equation}
\begin{array}{cccc}
    T^{\theta}_{~r\theta}\!\! &= \frac{1-e^{b(t,r)/2}}{r}, \quad &T^{t}_{~t r}\!\! &= -\frac{a'(t,r)}{2},   \\
    T^{\psi}_{~r\psi}\!\!     &= \frac{1-e^{b(t,r)/2}}{r}, \quad &T^{r}_{~t r}\!\! &= \frac{\dot b(t,r)}{2}.
\end{array}
\end{equation}
According to the transformation relations (\ref{trans}) of the tetrad fields, we can get
\begin{eqnarray}
\tilde T^{\rho}_{~\mu\nu} &=& T^{\rho}_{~\mu\nu} + [\Omega^{-1}\delta^{\rho}_{\nu}\partial_{\mu}\Omega
- \Omega^{-1}\delta^{\rho}_{\mu}\partial_{\nu}\Omega] \nonumber \\
&=& T^{\rho}_{~\mu\nu} + \bigg[\frac{\delta^{\rho}_{\nu}\partial_{\mu}\phi(t,r)}{2\phi(t,r)}
- \frac{\delta^{\rho}_{\mu}\partial_{\nu}\phi(t,r)}{2\phi(t,r)}\bigg]
\end{eqnarray}
Because $\partial_{\theta}\phi(t,r)=0$,
the additional items $\tilde T^{t}_{~\theta\mu}$ and $\tilde T^{r}_{~\theta\mu}$
both disappear in the $E_2^{~0}$ and $E_2^{~1}$,
and the result is not different from that we have proved.
The Birkhoff's Theorem in $f(T)$ gravity with off diagonal tetrad field in Einstein frame still holds.

Then we transform back the metric from Einstein frame to Jordan frame.
According to the transformation relations (\ref{trans}) of the tetrad fields,
we can get the metric in Jordan frame
\begin{eqnarray}
{\rm d} s^2
&=& \phi(t,r)^{-1}{\rm e}^{a\!(r)}~{\mathrm{d}}t^2-\phi(t,r)^{-1}{\rm e}^{b\!(r)}~{\mathrm{d}}r^2 \nonumber \\
&-& \phi(t,r)^{-1}r^2{\mathrm{d}}\theta^2-\phi(t,r)^{-1}r^2 \sin^2\!\theta~{\mathrm{d}}\phi^2.
\end{eqnarray}
Obviously, the metric in the Jordan frame clearly depends on time,
indicating that the Birkhoff's theorem is not satisfied.
This result suggests the non-physical equivalence between both frames.

On the other hand, the $E_1^{~0}$ and $E_0^{~2}$, which we introduce to prove the validity
of Birkhoff's Theorem with diagonal tetrad field, change as,
\begin{eqnarray}
E_1^{~0} &=& -\frac{{\rm e}^{-a(t,r)}}{2 r} \dot b(t,r) f_T
+ \frac{\tilde\partial^{\mu}\varphi}{2\sqrt{2\omega-3}}\tilde T^{t}_{~r\mu},\label{e10T} \\
E_0^{~2} &=& \frac{2{\rm e}^{- b(t,r)} \cos\theta}{r^4 \sin\theta} \bigg(\dot{a}^\prime(t,r) r - \dot{b}(t,r) \nonumber \\
&-& r \dot{b}(t,r) a^\prime(t,r)\bigg)\cdot f_{TT} + \frac{\tilde\partial^{\mu}\varphi}{2\sqrt{2\omega-3}}\tilde T^{\theta}_{~t\mu}.
\end{eqnarray}
Considering the diagonal tetrad field,
the non-zero components of the antisymmetric torsion tensor $T^{\rho}_{~\mu\nu}$ for the two covariant indices,
\begin{equation}
\begin{array}{cccc}
    T^{\theta}_{~r\theta}\!\!  &= \frac{1}{r}, \quad &T^{t}_{~t r}\!\! &= -\frac{a'(t,r)}{2},    \\
    T^{\psi}_{~r\psi}\!\!      &= \frac{1}{r}, \quad &T^{r}_{~t r}\!\! &= \frac{\dot b(t,r)}{2}, \\
    T^{\psi}_{~\theta\psi}\!\! &= \cot\!\theta .
\end{array}
\end{equation}
The additional item $\tilde\partial^{\theta}\varphi(t,r) \tilde T^{\theta}_{~t\theta}$ disappears
in the $E_0^{~2}$ for $\tilde\partial^{\theta}\varphi(t,r)=0$,
but the equation (\ref{e10T}) changes as
\begin{equation}
E_1^{~0} = -\frac{{\rm e}^{-a(t,r)}}{2 r} \dot b(t,r) f_T + \frac{\dot \varphi(t,r)}{2\sqrt{2\omega-3}}
\bigg(\frac{a'(t,r)}{2}+\frac{\phi'}{2\phi}\bigg)
\end{equation}
So the result is different from that case we have proved with diagonal tetrad field.
This means that $\dot \varphi(t,r)=0$ or $\dot \phi(t,r)=0$,
because that the non-diagonal elements of energy-momentum tensor are naturally equal to zero.
The Birkhoff's Theorem in $f(T)$ gravity with diagonal tetrad field
in Einstein frame still holds for $\phi=\phi(r)$.
Consequently, the metric in the Jordan frame clearly does not depend on time,
indicating that the Birkhoff's theorem is still satisfied.

In the above analysis, we have studied the equivalence between both Einstein frame and Jordan frame.
If we do not consider the ill-defined $\phi \rightarrow 0^+$, because of $\phi=e^{\varphi/\sqrt{2\omega-3}}$,
the transformation relations (\ref{trans}) of the tetrad fields
only depend on the concrete form of the $\phi$ field.
In other words, the equivalence between both frames depends on
the constraint on the $\phi$ field by the theoretical model.
We also can find the uncertainty of the $\phi$ field comes from the freedom of tetrad field,
The extra six degrees of freedom in the off diagonal tetrad
conceal the physical meaning of the $\phi$ field depending on time.
The physical reasons for the above consequence of Birkhoff's theorem in $f(T)$ gravity is
that the concrete form of the tetrad field is not determined.

\section{Discussions and Conclusions}
\label{sec4}

\begin{figure}[!h]
  \centering
  \subfigure[outside] {\includegraphics[scale=0.3]{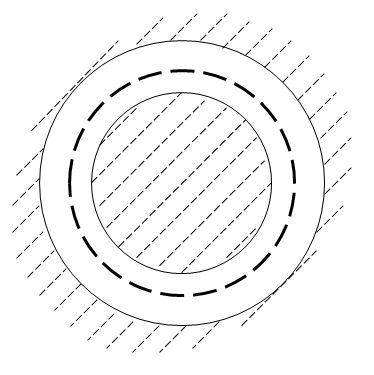}}
  \subfigure[inside] {\includegraphics[scale=0.5]{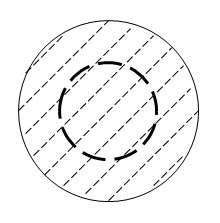}}
  \caption{the two cases of the Birkhoff's Theorem}
  \label{fig}
\end{figure}

In this letter we prove the validity of Birkhoff's theorem in $f(T)$ gravity with both the diagonal
and the off diagonal tetrad fields. In our previous work \cite{fT_birkhoff}, we have detailedly discussed
the physical meanings of the Birkhoff's theorem in ordinary conditions,
namely, the external vacuum gravitational field.
More generally, we consider a spherically symmetric matter distribution as shown in Fig.\ref{fig}(a).
The gravitational field of the interlining vacuum region is spherically symmetric,
because of the symmetric distribution of source matter. Accordingly,
the above analysis in this letter is applicable, and the gravitational field of the vacuum region is static.
The only property of the source matter may appear in $\widetilde a(r)$ of equation (\ref{A})
is the mass of internal source $M_I$ and mass of external source $M_E$.
(A similar problem is discussed specifically in \cite{ZhangSH,Penna} for black hole).
The radial motion and distribution of the source matter cannot affect the gravitational field any way.

The second conclusion is that the main feature of the Birkhoff's theorem is applicable in non-vacuum regions,
such as the case shown in Fig.\ref{fig}(b). Note that we actually do not claim that
the vanishing of density and pressure of matter is necessary to prove the validity of Birkhoff's theorem.
To obtain Eqs.(\ref{E20}, \ref{E21}), what we really demand is that the non-diagonal elements
of the energy-momentum tensor are zero, which is always satisfied for perfect fluid models.
As a conclusion, the gravitational field is static inside the spherically symmetric matter,
such as the region denoted by dashed line in Fig.\ref{fig}(b).
This conclusion is correct only if there is no radial motion or convection
across the sphere (the dashed line in Fig.\ref{fig}(b)).

As is known to all, Hawking's theorem \cite{Hawking} states that a stationary space-time
containing a black hole is a solution of the Brans-Dicke field equations with $V(\phi)=0$
if and only if it is a solution of the field equations,
and therefore it must be axially symmetric or static.
The proof of Hawking¡¯s theorem is performed in the Einstein frame,
in order to obtain that the rescaled Brans-Dicke scalar has canonical kinetic energy density
and obeys the weak and null energy conditions, and prove that the scalar field is static.
But we can find that an additional scalar-torsion coupling term
in the action (\ref{SE2}) breaks previous conditions.
And considering the scalar field equation (\ref{scalar field eqn0}) 
with $U(\varphi)$=$\frac{V(\phi)}{\phi^2}$=$0$ in vacuum,
which is different from $\tilde \Box \varphi=0$,
so we cannot vanish the contribution from the additional term on the horizon.
The Hawking's theorem in $f(T)$ gravity needs more detailed study.

In this letter we have proved Birkhoff's theorem validity in the $f(T)$ gravity with
both the diagonal and the off diagonal tetrad fields, and discussed the significations of this theorem both
in vacuum and non-vacuum conditions. We do not deal with perturbations in this present work,
which will be left for our future work in preparation.
The Birkhoff's theorem generally does not hold in the frame of $f(R)$ gravity by
using its scalar-tensor representation, and it is invalid at first linear order
of perturbations in the Jordan frames \cite{invalid fR2,fR_perturbation}.
Nevertheless, the case of $f(T)$ gravity is more complicated for an additional
scalar-torsion coupling term generated by conformal transformation \cite{fT_conformal}.
The perturbations in $f(T)$ gravity have been studied in \cite{pt}.
We can introduce some constraints on scalar field
like what S.~Capozziello, \textit{et al}, have done \cite{fR_perturbation},
by assuming a constant zero older scalar field as the background solution.
Consequently, the zero-order solution in perturbations will give the Schwarzschild-(Anti)de Sitter solution,
which will be proved in our next work \cite{fT_perturbation}.
We must make certain whether the higher order of the additional scalar-torsion and scalar field
would disappear like in the case of $f(R)$ gravity, or they still exist and have some effect, which may affect
the validity of Birkhoff's theorem in the perturbative approach.
We will deal with this problem in detail in our next work.

\section*{Acknowledgement}
We cordially thank Prof. Lewis H Ryder and Prof. Sergei D Odintsov for lots of interesting discussions
on possible roles the torsion may play in gravity and cosmology physics during the
project over years. This work is partly supported by Natural
Science Foundation of China under Grant Nos.11075078 and 10675062
and by the project of knowledge Innovation Program (PKIP) of Chinese
Academy of Sciences (CAS) under the grant No. KJCX2.YW.W10 through
the KITPC where we have initiated this present work.


\begin{thebibliography}{}
%
\bibitem{fT0} R.~Ferraro and F.~Fiorini, Phys.~Rev.~D \textbf{75}, 084031 (2007).
\bibitem{fT} G.~R.~Bengochea and R.~Ferraro, Phys.~Rev.~D \textbf{79}, 124019 (2009).
\bibitem{fT1} E. V. Linder, Phys. Rev. D \textbf{81}, 127301 (2010).
\bibitem{einstein} A.~Einstein, Sitzungsber. Preuss. Akad. Wiss. Phys. Math. Kl. 217 (1928); 224 (1928).
\bibitem{TEGR1} R.~Aldrovandi and J.~G.~Pereira, \textit{An Introduction to Teleparallel Gravity}
                Instituto de Fisica Teorica, UNSEP, Sao Paulo (http://www.ift.unesp.br/gcg/tele.pdf) (2007).
\bibitem{TEGR2} J. Garechi (2010), arXiv: 1010.2654.
\bibitem{fR0} S.~Nojiri, S.~D.~Odintsov, Gen.~Rel.~Grav.~\textbf{36}, 1765-1780 (2004).
\bibitem{fRrev} S.~Nojiri, S.~D.~Odintsov, Phys.~Rept.~\textbf{505}, 59-144 (2011).
\bibitem{fR1} S.~Nojiri and S. D. Odintsov, Int. J. Geom. Methods Mod. Phys. \textbf{4}, 115 (2007).
\bibitem{fR11} Miao Li \textit{et al}, Commmun. Theor. Phys. \textbf{56}, 525-604 (2011).
\bibitem{fRa0} X.H.Meng and P.Wang, Class. Quant. Grav. \textbf{20}, 4949 (2003).
\bibitem{fRa1} X.H.Meng and P.Wang, Class. Quant. Grav. \textbf{21}, 951 (2004).
\bibitem{fRa2} X.H.Meng and P.Wang, Class. Quant. Grav. \textbf{21}, 2029 (2004).
\bibitem{fRa3} X.H.Meng and P.Wang, Class. Quant. Grav. \textbf{22}, 23 (2005).
\bibitem{fRa4} X.H.Meng and P.Wang, Gen. Rel. Grav. \textbf{36}, 1947 (2004).
\bibitem{fRa5} X.H.Meng and P.Wang, Phys. Lett. B \textbf{584}, 1 (2004);
               P.Wang and X.H.Meng, Class. Quant. Grav. \textbf{22} 283 (2005);
               J.Ren and X.H.Meng, Phys. Lett. B \textbf{633} 1 (2006); ibid, \textbf{636}, 5 (2006);
               M.G.Hu and X.H.Meng, ibid, \textbf{635} 186 (2006);
               J.Ren and X.H.Meng, Int'l Jour.Mod.Phys.D16  1341(2007).
\bibitem{fRa6} E.Flanagan, Class. Quant. Grav. \textbf{21}, 417 (2003).
\bibitem{fRa7} S. Nojiri and S. Odintsov, Phys. Lett. B, \textbf{576}, 5 (2003).
\bibitem{fRa8} S. Nojiri and S. Odintsov, Phys. Rev. D \textbf{68}, 123512 (2003).
\bibitem{fRa9} D.Volink, Phys. Rev. D \textbf{68} 063510 (2003).
\bibitem{fR2} S.~Capozziello and M.~Francaviglia,~Gen.~Relativ.~Gravit. \textbf{40}, 357 (2008).
\bibitem{fR3} T.~Sotiriou and V.~Faraoni, Rev. Mod. Phys. \textbf{82}, 451-497 (2010).
\bibitem{fR4} A. De Felice and S. Tsujikawa, Living Rev. Relativity \textbf{13}, 3 (2010).
\bibitem{fT_Lorentz} B. J. Li, T. P. Sotiriou and J. D. Barrow, Phys. Rev. D \textbf{83}, 064035 (2011).
\bibitem{fT_conformal} R.~J. Yang, Euro. Phys. Lett. \textbf{93}, 60001 (2011).
\bibitem{fT_w} Kazuharu Bamba \textit{et al}, JCAP \textbf{1101}, 021 (2011);
               Kazuharu Bamba \textit{et al}, (2010) arXiv:1008.4036.
\bibitem{fT2} P. Wu and H. Yu, Phys.~Lett.~B \textbf{693}, 415-420 (2010).
\bibitem{fT3} R. Myrzakulov (2010), arXiv: 1006.1120.
\bibitem{fT4} K.K.Yerzhanov \textit{et al}, (2010), arXiv: 1006.3879.
\bibitem{fT5} R.~J.~Yang Eur. Phys. J. C \textbf{71}, 1797 (2011).
\bibitem{fT6} G.~R.~Bengochea, Phys. Lett. B \textbf{695}, 405-411 (2011).
\bibitem{fT7} M. Hamani Daouda \textit{et al}, Euro. Phys. J. C \textbf{71}, 1817 (2011);
              M. Hamani Daouda \textit{et al}, Euro. Phys. J. C \textbf{72}, 1890 (2012).
\bibitem{fT_birkhoff} Xin-he Meng, Ying-bin Wang, Euro. Phys. Lett. \textbf{71}, 1755 (2011).
\bibitem{birkhoff} G.~D.~Birkhoff, \textit{Relativity and Modern Physics}
                   (Harvard University Press, Cambridge, 1923) 4, 5, 11.
\bibitem{bb} S.~Deser, J.~Franklin, Am.~J.~Phys. \textbf{73}, 261 (2005);
             N.~V.~Johansen, F.~Ravndal, Gen.~Relat.~Grav. \textbf{38}, 537 (2006);
             S.~Deser, Gen.~Relat.~Grav. \textbf{37}, 2251 (2005).
\bibitem{weinberg} S.~Weinberg, \textit{Gravitation and Cosmology} (Wiley, New York, 1972).
\bibitem{fT_relativistic stars} C.~G.~B\"ohmer, A.~Mussa and N.~Tamanini (2011), arXiv:1107.4455v2.
\bibitem{invalid fR2} S.~Capozziello, A.~Stabile and A.~Troisi, Phys. Rev. D \textbf{76}, 104019 (2007).
\bibitem{invalid fR3} V.~Faraoni, Phys. Rev. D \textbf{81}, 044002 (2010).
\bibitem{ZhangSH} Y.~Liu, S.~N.~Zhang (2009), Phys. Lett. B \text(679), 88-94 (2009).
\bibitem{Penna} Robert F.~Penna, Phys. Lett. B \textbf{707}, 233-236 (2012).
\bibitem{Hawking} S.~W.~Hawking, Comm. Math. Phys. \textbf{25}, 167 (1972);
                  S.~W.~Hawking, Comm. Math. Phys. \textbf{25}, 152 (1972).
\bibitem{fR_perturbation} S.~Capozziello and D.~S\'{a}ez-G\'{o}mez (2011), arXiv:1107.0948.
\bibitem{pt} S.~H.~Chen \textit{et al}, Phys.~Rev.~D \textbf{83}, 023508 (2011);
             J.~B.~Dent, S.~Dutta, E.~N.~Saridakis, JCAP \textbf{1101}, 009 (2011);
             Yi-Fu Cai \textit{et al}, Class. Quantum. Grav. \textbf{28}, 215011 (2011).
\bibitem{fT_perturbation} Han Dong, Ying-bin Wang and Xin-he Meng,
                 \textit{The validity of Birkhoff's Theorem in the $f(T)$ Gravity using a perturbative approach}

%
\end{thebibliography}
%

%
\end{document}